\documentclass[trackchanges]{aastex7}
\usepackage{amsmath}
\usepackage{subcaption}

\begin{document}

\title{The Photochemical Plausibility of Warm Exo-Titans Orbiting M-Dwarf Stars}

\correspondingauthor{Sukrit Ranjan}
\email{sukrit@arizona.edu}

\author[orcid=0000-0002-5147-9053,sname='Ranjan']{Sukrit Ranjan}
\altaffiliation{Blue Marble Space Institute of Science}
\affiliation{Lunar \& Planetary Laboratory, University of Arizona, Tucson, AZ 85719}
\email{sukrit@arizona.edu}

\author[0000-0002-0413-3308,sname=Wogan]{Nicholas F. Wogan} 
\altaffiliation{NPP Fellow}
\affiliation{NASA Ames Research Center, Moffett Field, CA 94035}
\email{nicholas.f.wogan@nasa.gov}

\author[0000-0002-5322-2315, sname=Glidden]{Ana Glidden}
\affiliation{Department of Earth, Atmospheric and Planetary Sciences, Massachusetts Institute of Technology, Cambridge, MA 02139, USA}
\email{aglidden@mit.edu}

\author[0000-0001-5128-585X,sname=Wang]{Jingyu Wang}
\affiliation{Lunar \& Planetary Laboratory, University of Arizona, Tucson, AZ 85719}
\email{wangjingyu@arizona.edu}

\author[0000-0002-7352-7941]{Kevin B. Stevenson}
\affiliation{Johns Hopkins APL, 11100 Johns Hopkins Rd, Laurel, MD 20723, USA}
\affiliation{Consortium on Habitability and Atmospheres of M-dwarf Planets (CHAMPs), Laurel, MD, USA}
\email{Kevin.Stevenson@jhuapl.edu}

\author[0000-0002-8507-1304, sname=Lewis]{Nikole Lewis}
\affiliation{Cornell University, Department of Astronomy, Ithaca, NY 14850}
\email{nikole.lewis@cornell.edu}



\author[0000-0003-3071-8358, sname=Koskinen]{Tommi Koskinen}
\affiliation{Lunar \& Planetary Laboratory, University of Arizona, Tucson, AZ 85719}
\email{tommik@arizona.edu}

\author[0000-0002-6892-6948]{Sara Seager}
\affiliation{Department of Earth, Atmospheric and Planetary Sciences, Massachusetts Institute of Technology, Cambridge, MA 02139, USA}
\affiliation{Department of Physics and Kavli Institute for Astrophysics and Space Research, Massachusetts Institute of Technology, Cambridge, MA 02139, USA}
\affiliation{Department of Aeronautics and Astronautics, MIT, 77 Massachusetts Avenue, Cambridge, MA 02139, USA}
\email{seager@mit.edu}

\author[0000-0003-4328-3867]{Hannah R. Wakeford}
\affiliation{School of Physics, University of Bristol, HH Wills Physics Laboratory, Tyndall Avenue, Bristol BS8 1TL, UK}
\email{hannah.wakeford@bristol.ac.uk}

\author[0000-0001-7827-7825]{Roeland P. van der Marel}
\affiliation{Space Telescope Science Institute, 3700 San Martin Drive, Baltimore, MD 21218, USA}
\affiliation{Center for Astrophysical Sciences, The William H. Miller III Department of Physics \& Astronomy, Johns Hopkins University, Baltimore, MD 21218, USA}
\email{marel@stsci.edu}


\begin{abstract}
The James Webb Space Telescope (JWST) has begun to spectrally characterize small exoplanets orbiting M-dwarf stars, but interpretation of these spectra is ambiguous, with stellar, instrumental, or atmospheric origins possible for apparent spectral features. Consequently, interpretation of JWST small exoplanet spectra follows a Bayesian approach, with less theoretically plausible interpretations facing a higher burden of proof. Here, we use photochemical modeling to evaluate the plausibility of warm exo-Titans, exoplanets with N$_2$-CH$_4$ atmospheres analogous to Titan but orbiting closer to their host stars. Consideration of warm exo-Titans is motivated by arguments from planet formation, as well as tentative evidence from observations. Using TRAPPIST-1e as a case study, we show that the higher instellation experienced by warm exo-Titans reduces their CH$_4$ lifetime $\tau_{\text{CH}_{4}}$ relative to true Titan by orders of magnitude, reducing the probability of observing them. We constrain the $\tau_{\text{CH}_{4}}$ on a warm exo-Titan to be $\leq0.1\times$ (and most likely $\leq0.02\times$) true Titan, implying the absolute probability of detecting a warm exo-Titan is $<0.1$ and likely $<0.01$. This finding is consistent with recent JWST nondetections of CH$_4$-dominated atmospheres on warm terrestrial exoplanets. The low prior probability means that the standard of proof required to claim a warm exo-Titan detection is high, and we offer specific suggestions towards such a standard of proof. Observation of oxidized carbon species would corroborate a putative warm exo-Titan detection. Confirmed detection of warm exo-Titans would signal the need to fundamentally rethink our understanding of the structure, dynamics, and photochemistry of Titan-like worlds. 
\end{abstract}

\keywords{\uat{Exoplanet atmospheric composition}{2021} --- \uat{Titan}{2186} --- \uat{Ocean planets}{1151} --- \uat{Extrasolar rocky planets}{511} --- \uat{Theoretical models}{2107} --- \uat{Methane}{1042} --- \uat{James Webb Space Telescope}{2291}}


\section{Introduction \label{sec:intro}} 
The James Webb Space Telescope (JWST) enables observational study of terrestrial exoplanet atmospheres. JWST thermal eclipse observations constrain the presence of thick atmospheres on hot, uninhabitable planets orbiting close to their stars (e.g., \citealt{Koll2019, Koll2022, Zieba2023, Hu2024}). Complementarily, JWST transmission spectroscopy can constrain the presence and composition of terrestrial exoplanet atmospheres (e.g., \citealt{Lustig-Yaeger2023, Damiano2024, Cadieux2024, Gressier2024}). However, in attempting to characterize terrestrial exoplanet atmospheres, JWST is operating at its limits. In many cases, numerous transits ($\geq$ 10) are required to build up adequate signal-to-noise (SNR), the expected amplitude of the signals is close to the instrument noise floors, and noise from stellar variability and heterogeneity additionally frustrates observations \citep{Rackham2018, Lustig-Yaeger2019, Zhan2021, Moran2023, Espinoza2025}. Stellar contamination is particularly problematic for terrestrial exoplanets as those accessible to characterization often orbit M-dwarf stars which are especially active \citep{Rackham2023, Seager2024}. 

These limitations mean that interpretation of apparent signals in JWST transmission spectra of terrestrial exoplanets is ambiguous. A putative exoplanet spectral feature can have instrumental, stellar, or atmospheric origins. Stellar contamination is a particularly acute problem because it is not yet well characterized in models \citep{Rackham2018, Rackham2023, Seager2024}, meaning that it is difficult to rule out as an explanation for the spectral features expected from terrestrial exoplanets (e.g., \citealt{Moran2023, Espinoza2025}). Consequently, interpretation of JWST small-planet spectra in practice follows a Bayesian approach, where the plausibility of an atmospheric scenario influences the confidence with which it is inferred from a spectrum \citep{Wogan2024a, Welbanks2025}. This follows practice in solar system science; for example, a higher standard of proof is demanded for the claim of seasonally-varying methane on Mars, because seasonally-varying Martian methane is difficult to reconcile with theory \citep{Zahnle2011, Yung2018, Viscardy2025}. Consequently, theoretical modeling is important to support interpretation of small planet spectra.  

In this paper, we use theory to evaluate the plausibility of warm exo-Titans, i.e., exoplanets analogous to Titan but at higher instellation (closer orbit) compared to Titan. Specifically, we define warm exo-Titans to be planets with N$_2$-CH$_4$ atmospheres with $T_{surf}\geq 200$ K\footnote{This choice of temperature is motivated by the threshold at which the volume mixing ratio of H$_2$O at the surface exceeds 1 ppm, a critical threshold in the atmospheric chemistry of Titan-like planets \citep{Zahnle2020}} overlying a planet composed primarily of rock and ice, with volatile mass fractions $>10\%$. Consideration of warm exo-Titans is motivated by arguments from planet formation and demographics that $\gtrsim10\%$ of close-orbiting super-Earths are ``water worlds" with $>10\%$ H$_2$O mass fraction \citep{Zeng2019, Luque2022, Chakrabarty2024}, which may consequently store and outgas copious CH$_4$ in their ice mantles \citep{Levi2014} as on Titan \citep{Tobie2012}. These facts have motivated theoretical \citep{Morley2017, Adams2022} and more recently observational (e.g., TRAPPIST-1e, \citealt{Glidden2025}; L98-59c, \citealt{Scarsdale2024}; LHS 475b, \citealt{Lustig-Yaeger2023}) consideration of the possibility of Titan-analog atmospheres on warm super-Earths. Additionally, warm Titans constitute potential abiotic false positives for CH$_4$ as a biosignature gas \citep{Thompson2022}. Consequently, it is relevant to explore the plausibility of warm exo-Titans.

We use a 1D atmospheric photochemical model to constrain the plausibility of the warm exo-Titan planetary scenario, relative to true Titan. We focus on the endmember of TRAPPIST-1e as an exo-Titan, motivated by the initial tentative (statistically insignificant) hints that it has a Titan-like atmosphere \citep{Espinoza2025, Glidden2025}; however, our basic findings generalize to M-dwarf warm exo-Titans as a class. The application of this work is to determine the standard of proof required to claim a warm exo-Titan: if exo-Titans are substantially less plausible than Titan, then a high standard of proof is required to claim the existence of one. We consider the implications of our findings for JWST observations \citep{Lustig-Yaeger2023, Scarsdale2024, Glidden2025}. 

\section{Background} \label{sec:background}
Investigations of exo-Titans (warm and cold) have focused on their predicted atmospheric observables with less emphasis on their plausibility. Starting with Titan-like bulk atmospheric compositions, several works have predicted the resulting climate states, trace gas compositions, and spectral observables \citep{Lora2018, Adams2022, Mandt2022}. However, less work has evaluated the plausibility of the Titan-like bulk atmospheric composition to begin with. CH$_4$ on Titan is photochemically unstable, with a lifetime of $10^7-10^8$ years \citep{Yung1984, Wilson2004, Atreya2006, Nixon2018}. Isotopic evidence suggests that the carbon in Titan's atmosphere is geologically recent ($<940$ Ma; \citealt{Mandt2012}), while the presence of $^{40}$Ar is indicative of outgassing \citep{Tobie2014}. Taken together, these facts suggest Titan's CH$_4$-rich atmosphere is a transient oddity attributable to a geologically recent outgassing event \citep{Tobie2006}, and that in steady-state Titan may lack a CH$_4$-rich atmosphere \citep{Wong2015}. This means that we may have ``gotten lucky" in seeing a CH$_4$-rich Titan today, and conversely that detecting a CH$_4$-rich exo-Titan is statistically unlikely. The lifetime problem is exacerbated when we consider warm exo-Titans, which are subject to higher levels of UV flux compared to Titan and which may naively be expected to feature even shorter CH$_4$ lifetimes.

\citet{Turbet2018} were the first to consider the plausibility of warm exo-Titans, through the case study of the TRAPPIST-1 system. They argued that it was unlikely for the TRAPPIST-1 planets to be exo-Titans because CH$_4$ lifetimes are 1-3 orders of magnitude shorter compared to Titan, assuming the CH$_4$ lifetime scales inversely with the EUV flux. However, they emphasized that their calculation neglected detailed photochemistry, in particular the possibility of Titan-like haze formation that might inhibit CH$_4$ photolysis \citep{Arney2016}, increasing the CH$_4$ lifetime. \citet{Thompson2022} estimated methane lifetimes on warm exo-Titans assuming the lifetime to be set by diffusion-limited hydrogen escape, finding a lifetime for all planetary CH$_4$ to be 10 Myr per planet percent mass fraction H$_2$O assuming Titan's CH$_4$:H$_2$O ratio. For a Titan-like H$_2$O mass fraction of $\sim50\%$ \citep{Tobie2012, CatlingKasting2017}, this would imply a CH$_4$ lifetime of $\sim500$ Myr, which begins to approach geologically long-lived. However, this approach implicitly neglects the possibility that CH$_4$ loss is photolysis-limited, as well as the possibility that CH$_4$ may be lost at a rate exceeding the hydrogen production/loss rate, due to e.g., reactions with carbon radicals and subsequent sedimentation as aerosol. In summary, previous work has estimated the CH$_4$ lifetime on warm exo-Titans but have not resolved the detailed photochemistry (i.e., haze formation, temperature-dependent reaction rate changes, lifetime estimates) relevant to the problem. Here, we undertake a detailed photochemical analysis of the likelihood of observing a warm exo-Titan around an M dwarf star like TRAPPIST-1.

\section{Methods} \label{sec:methods}

\subsection{Photochemical Model \& Configuration}
We use the \texttt{Photochem} photochemical model \citep{Wogan2024_photochem} to estimate CH$_4$ lifetimes on warm exo-Titans (Section~\ref{sec:ch4_lifetime}). This model is an appropriate tool for this study because it is a general exoplanet photochemistry model validated against Titan \citep{Wogan2023}. Importantly, \texttt{Photochem} resolves organic haze production which is identified as a process that may affect CH$_4$ lifetimes \citep{Turbet2018}, and its CH$_4$ photochemical network has recently been intercompared with other photochemical models \citep{Wogan2024a}, making it appropriate for this study. \texttt{Photochem}'s neglect of ion chemistry is justified because the CH$_4$ lifetime is controlled by neutral chemistry \citep{Yung1984}, as demonstrated by the insensitivity of CH$_4$ lifetime estimates to inclusion of ion chemistry \citep{Yung1984, Wilson2004}. We use the chemical network shipped with \texttt{Photochem}, and consider species composed of H, N, O and C. We exclude S and Cl-bearing species by analogy to Titan, and because their volcanogenic emission to the atmosphere on a water world is unlikely \citep{Kite2009, Krissansen-Totton2021waterworld}. 

We use \texttt{Photochem} to model TRAPPIST-1e as an exo-Titan. We adopt mostly closed-box boundary conditions (0 wet, dry deposition; 0 escape) as implemented in its Titan template; here, we detail the exceptions to this rule. We adopt a fixed surface-pressure boundary condition of pN$_2=1.5$ bar, motivated by Titan. We adopt a fixed surface-pressure boundary condition for CH$_4$ ranging from p$_{\text{CH}_{4}}=3\times10^{-7}$ bar to $0.15$ bar, to explore the effect of p$_{\text{CH}_{4}}$ on the methane lifetime ($\tau_{\text{CH}_{4}}$) since a wide range of p$_{\text{CH}_{4}}$ is considered for exoplanets \citep{Lustig-Yaeger2023, Glidden2025}. We do not consider p$_{\text{CH}_{4}}>0.15$ bar because climate modeling shows that for p$_{\text{CH}_{4}}\gtrapprox 0.1-1$ bar, the surface temperature increases rapidly with p$_{\text{CH}_{4}}$ \citep{Turbet2018}, possibly melting the surface and liberating more CH$_4$ in a positive feedback loop \citep{Thompson2022}; more detailed planetary evolution modeling including climate feedback is required to explore this regime. Below this threshold, increasing p$_{\text{CH}_{4}}$ is predicted to cool the planet \citep{Turbet2018}, which constitutes a negative feedback loop that should stabilize p$_{\text{CH}_{4}}$; we focus on this regime. We assume abundant surface H$_2$O, as water at the substellar point and as ice elsewhere \citep{Sergeev2022}, and therefore adopt a fixed surface pressure boundary condition for H$_2$O of pH$_2$O=$2.7\times10^{-4}$ bar, corresponding to the saturation pressure of water over ice \citep{Murphy2005} at the adopted surface temperature of 240 K (see below). 

We permit H and H$_2$ to escape at the diffusion-limited rate. We adopt deposition velocities for CO and O$_2$ of $10^{-9}$ cm s$^{-1}$, corresponding to the deposition velocities estimated by \citet{Harman2015} for an abiotic ocean scaled by $0.1$ to reflect the likely smaller ocean coverage fraction. We adopt a deposition velocity for CO$_2$ of $10^{-7}$ cm s$^{-1}$, corresponding to the modern Earth deposition velocity \citep{Hu2012} scaled by $0.1$ to reflect the likely smaller ocean coverage fraction and by a further $0.01$ to reflect presumably less efficient carbonate formation on cold TRAPPIST-1e. While these values are highly uncertain, it is advisable to prescribe nonzero deposition velocities for the oxidized carbon species because they can be formed by H$_2$O photooxidation of CH$_4$ in the warm Titan scenario (Section~\ref{sec:ch4_lifetime}). We choose a haze particle radius of $r_d=1~\mu$m. We adopt surface deposition velocities for haze-forming species (C$_2$H$_6$, C$_2$H$_4$, C$_2$H$_2$, CH$_3$CN, HC$_3$N, HCN) of $10^{-5}$ cm/s to mitigate the risk that our haze formation mechanism is incomplete \citep{Hu2012}. 

\begin{figure}[h]
\centering
\begin{subfigure}{0.45\linewidth}
    \includegraphics[width=1\linewidth]{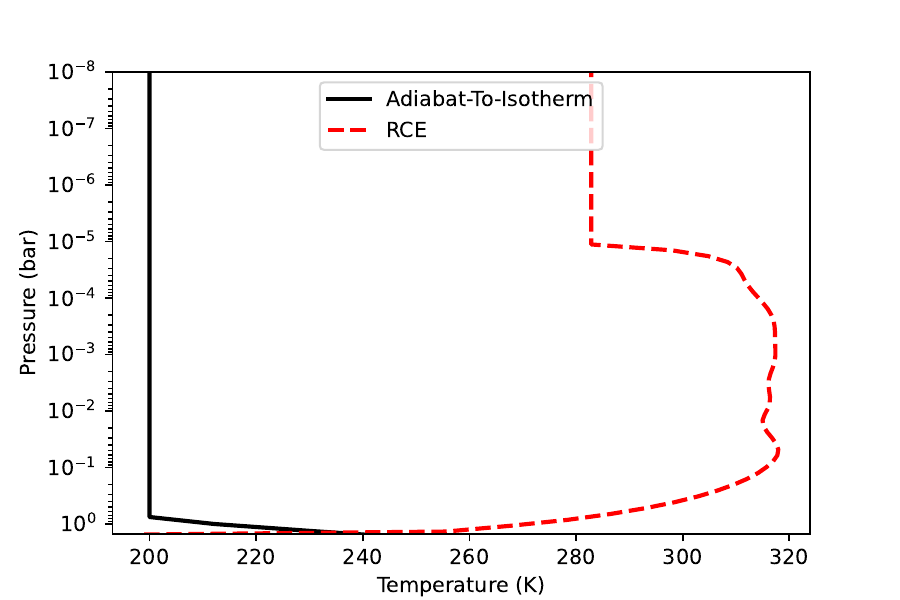}
\caption{Temperature-pressure profiles considered in this work. The black line shows our baseline $P_{surf}=1.5$ bar, $T_{surf}=240$ K T-P profile, calculated assuming adiabatic evolution to an isothermal stratosphere. The red dashed line shows the radiative-convective equilibrium (RCE) temperature-pressure profile calculated by \texttt{Photochem}, considered as a sensitivity test. } 
\label{fig:TP}
\end{subfigure}
\hfill
\begin{subfigure}{0.5\linewidth}
    \includegraphics[width=1\linewidth]{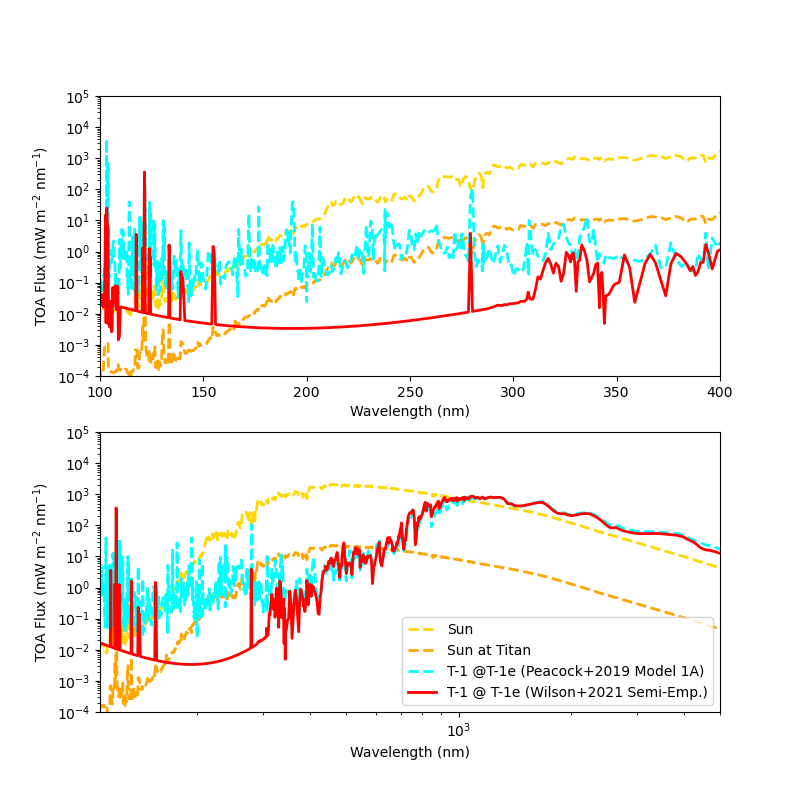}
    \caption{SEDs adopted in this work. The \citet{Wilson2021} semi-empirical SED (red solid line) is the baseline SED used to force our model to simulate TRAPPIST-1e in our calculations. For our model validation and for our sensitivity tests (shown with dashed lines), we also considered solar instellation as well as the \citet{Peacock2019a} model 1A TRAPPIST-1 SED.} 
\label{fig:SED}
\end{subfigure}
\caption{Key photochemical modeling inputs.}
\end{figure}

We force our model with the semi-empirical TRAPPIST-1 spectral energy distribution (SED; 1 nm - 100 $\mu$m) of \citet{Wilson2021}. As the SED of TRAPPIST-1 is highly uncertain, we sensitivity test our calculations to the SED of \citet{Peacock2019a} (their model 1A; Table~\ref{tbl:senstests}). We assume a surface temperature of 240K, motivated by the surface temperatures predicted for N$_2$-CH$_4$ atmospheres on TRAPPIST-1e for p$_{\text{CH}_{4}}\leq1$ bar \citep{Turbet2018}, evolving to an isothermal stratosphere at 200K following \citet{Hu2012}. While a common approach (e.g., \citealt{Kasting1988, Aguichine2021, Wogan2023}), this approximation may be inaccurate for CH$_4$-dominated atmospheres orbiting M-dwarf stars, for which antigreenhouse effects are anticipated \citep{RamirezKaltenegger2018, Turbet2018}. Antigreenhouse effects occur when an atmospheric gas significantly absorbs incoming stellar radiation, which heats the upper atmosphere and therefore by energy balance cools the surface \citep{PPC}. Indeed, a simulation with \texttt{Photochem}'s climate module indicated very strong antigreenhouse effect with surface temperatures of 200K and stratospheric temperatures approaching 320K for a planet with p$_{\text{CH}_{4}}=0.03$ bar (Figure~\ref{fig:TP}). This is a sufficiently extreme result that we are disinclined to rely on it without further verification from independent climate models. We therefore use the adiabat-to-isotherm as our base case, and also perform calculations with the \texttt{Photochem}-calculated radiative-convective equilibrium (RCE) temperature-pressure profile in Figure \ref{fig:TP} as a sensitivity test (Table~\ref{tbl:senstests}). We assume a vertical eddy diffusion coefficient, $K_{ZZ}=10^6$ cm$^{2}$ s$^{-1}$ throughout the atmosphere, an intermediate value for the Solar System terrestrial planets \citep{ZhangShowman2018}. 

We run the model to an integration time of $10^{16}$ s. To confirm this is an adequate convergence criteria, we evolved each simulation in Figure \ref{fig:LCH4_PCH4} from $10^{16}$ to $10^{17}$ s, and visually confirmed that the main atmospheric species (e.g., CO, CO$_2$, H$_2$, etc.) do not change in concentration. To further verify steady-state convergence, we have confirmed that our simulations balance flux to within a factor of $\sim10^{-5}$ or better. For example, in the p$_{\text{CH}_{4}}=3\times10^{-2}$ bar simulation of Figure \ref{fig:LCH4_PCH4}, the column integrated net chemical production of H$_2$ is $2.5999546 \times 10^{11}$ cm$^{-2}$ s$^{-1}$ while the escape to space is $-2.5999543 \times 10^{11}$ cm$^{-2}$ s$^{-1}$, so the H$_2$ fluxes balance to a factor of $\sim10^{-7}$.





As a test of our approach, we ran {\tt Photochem} on the Titan validation case of \citet{Wogan2023} and computed $\tau_{\text{CH}_{4}}=\frac{N_{\text{CH}_{4}}}{L_{\text{CH}_{4}}}$, where $N_{\text{CH}_{4}}$ is the CH$_4$ column number density and $L_{\text{CH}_{4}}$ is the column-integrated CH$_4$ net chemical loss rate. We found $\tau_{\text{CH}_{4}}=2\times10^{7}$ years, within the $10^{7}-10^{8}$ year range estimated for Titan \citep{Yung1984, Wilson2004, Atreya2006, Nixon2018}. 

\subsection{Simulated Spectra} \label{sec:forwardmodels_retrievals}

For a few of our simulated atmospheres (Figure~\ref{fig:COx}), we compute simulated transmission spectra to evaluate the detectability of photochemically produced species on warm exo-Titans (Section~\ref{sec:cox}). To do this, we use the radiative transfer package {\tt petitRADTRANS} ({\tt pRT}; \citealt{Molliere2019prt}). We include molecular absorption from CH$_4$ \citep{Yurchenko2017}, CO$_2$ \citep{Yurchenko2020}, H$_2$O \citep{Rothman2010}, CO \citep{Rothman2010}, C$_2$H$_2$ \citep{Chubb2020}, C$_2$H$_4$ \citep{Mant2018}, HCN \citep{Barber2014}, NH$_3$ \citep{Coles2019}, O$_2$ \citep{Gordon2017}, O$_3$ \citep{Rothman2013}. We construct an equally log-spaced grid in pressure ranging from a surface of $\sim$1.5 bars, reminiscent of the surface of Titan to very low, high altitude pressures of $\sim10^{-10}$. We include the effects of Rayleigh scattering from N$_2$, O$_2$, and H$_2$ and collision-induced absorption (CIA) from N$_2$-N$_2$, O$_2$-O$_2$, N$_2$-O$_2$, N$_2$-H$_2$, H$_2$-H$_2$ \citet[][and references therein]{DalgarnoWilliams1962, Borysow1988, Borysow1989, Richard2012, Thalman2014, Thalman2017, Molliere2019prt}. We do not introduce clouds into our forward models, but use a power-law model to describe the haze, $$\kappa=\kappa_0\bigg(\frac{\lambda}{\lambda_0}\Bigg)^{\gamma}$$where $\kappa$ is the opacity of the scattering-cross section. We do not adopt a Rayleigh scattering slope of $\gamma=-4.0$, but adopt the value of $-1.9$ fit by \citet{Robinson2014b} to Titan's transmission spectrum. We use {\tt PandExo} \citep{Batalha2017PANDEXO} to bin the data to the wavelength range and resolution of JWST NIRSpec PRISM (0.6-5.3$\mu$m; R$\approx$100). In Figure \ref{fig:COx}, we show the results of our simulated transmission spectra for the \citet{Wilson2021} SED (b) and the \citet{Peacock2019a} SED (c). The molecules that are the strongest sources of the opacity are shown as shaded colors. In both cases, CH$_4$ spectral features dominate followed by CO. Notably, CO$_2$ is only seen for the \citet{Peacock2019a} SED. Additionally, the CO feature is much stronger for the \citet{Peacock2019a} SED. 




\section{Results \& Discussion} \label{sec:results_discussion}

\subsection{Methane Lifetime \label{sec:ch4_lifetime}}
\begin{figure}[h]
\centering
\includegraphics[width=0.8\linewidth]{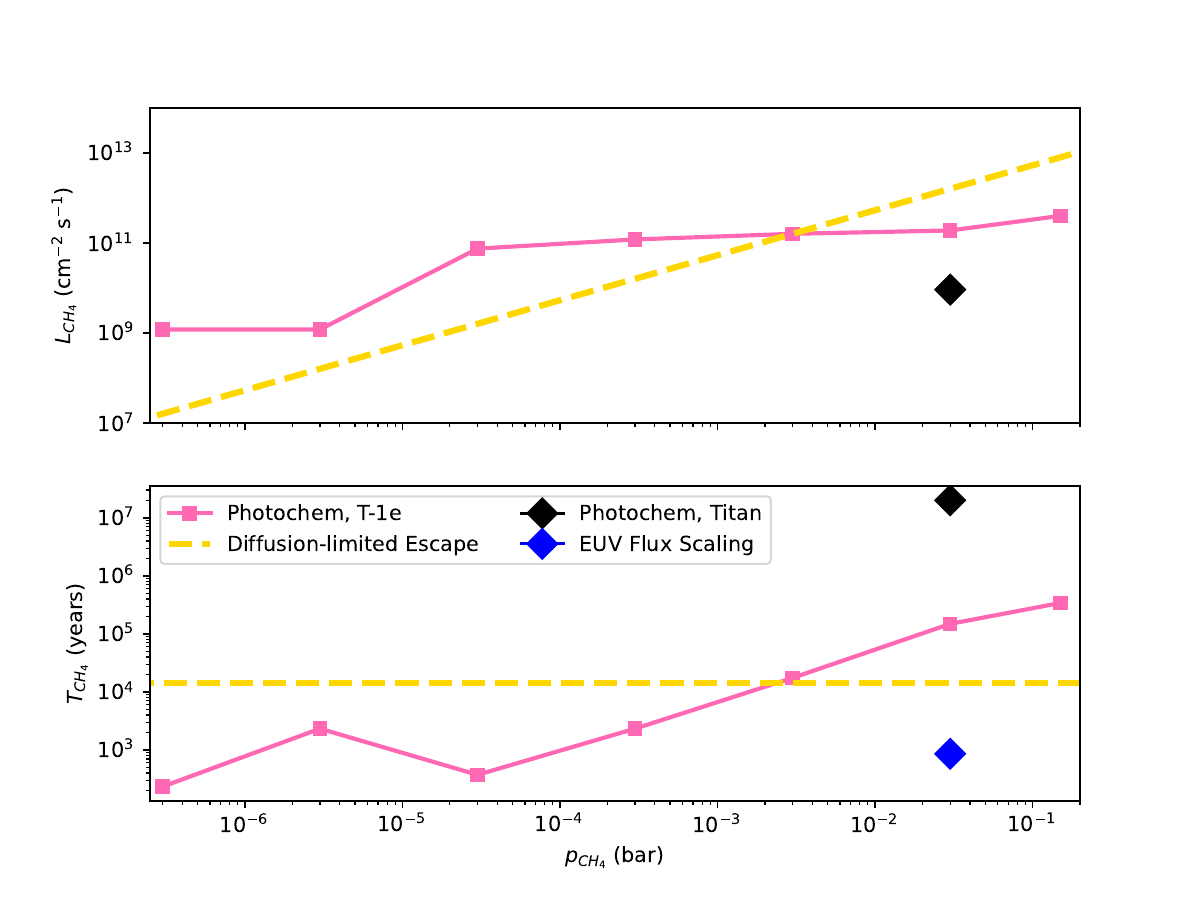}
\caption{CH$_4$ net chemical loss rate ($L_{\text{CH}_{4}}$) and lifetime ($\tau_{\text{CH}_{4}}$) as a function of the partial surface pressure of methane. The pink line corresponds to \texttt{Photochem} calculations. The yellow line corresponds to CH$_4$ loss estimated via diffusion-limited escape of equivalent hydrogen, following \citealt{Thompson2022}. The black diamond corresponds to true Titan. The blue diamond corresponds to the assumption that $\tau_{\text{CH}_{4}}\propto \frac{1}{F_{EUV}}$ \citep{Turbet2018}. The CH$_4$ loss rate increases sublinearly with CH$_4$ concentration, meaning that the lifetime increases with p$_{\text{CH}_{4}}$ and exceeds both the diffusion-limited hydrogen escape estimate and the $F_{EUV}$ estimate at high p$_{\text{CH}_{4}}$. However, the TRAPPIST-1e CH$_4$ lifetime remains orders of magnitude below the true Titan CH$_4$ lifetime across the range of p$_{\text{CH}_{4}}$ considered, meaning that a warm M-dwarf exo-Titan is much less likely than a true Titan.} 
\label{fig:LCH4_PCH4}
\end{figure}

We find $\tau_{\text{CH}_{4}}$ on a warm exo-Titan orbiting TRAPPIST-1e to be very short compared to true Titan (Figure~\ref{fig:LCH4_PCH4}). For a Titan-like p$_{\text{CH}_{4}}=0.03$ bar, $\tau_{\text{CH}_{4}}=2\times10^{5}$ years for our warm exo-Titan, compared to $2\times10^{7}$ years estimated for true Titan.  Despite consideration of haze formation, $\tau_{\text{CH}_{4}}|_{\text{warm exo-Titan T-1e}}\leq1\times10^{-2}\times\tau_{\text{CH}_{4}}|_{\text{true Titan}}$, meaning that a warm exo-Titan TRAPPIST-1e is much less plausible than true Titan. This basic finding is robust to sensitivity tests to the assumed photochemical parameters, with the strongest sensitivity to choice of SED (Table~\ref{tbl:senstests}). To additionally confirm the controlling role of the SED, we computed  $\tau_{\text{CH}_{4}}$ for our baseline scenario except forced by the Solar spectrum at 9.5 AU and found $\tau_{\text{CH}_{4}}=5\times10^{6}$ years, within a factor of 4 of true Titan. This shows that of the factor of 100 decrease in  $\tau_{\text{CH}_{4}}$, a factor of 25 is due to the SED and a factor of 4 is due to all other changes in the planet scenario ($M_P$, $R_P$, T-P profile, boundary conditions), reinforcing the conclusion that the SED is the main control on $\tau_{\text{CH}_{4}}$.

\begin{deluxetable}{p{14 cm}p{2 cm}}
    \tabletypesize{\footnotesize}
    \tablecolumns{2} 
    \tablecaption{Sensitivity of $\tau_{\text{CH}_{4}}$ (p$_{\text{CH}_{4}}=0.03~\text{bar})$ for TRAPPIST-1e as a warm exo-Titan to the assumed photochemical parameters.   \label{tbl:senstests}}
    \tablehead{Condition & $\tau_{\text{CH}_{4}}$ (years)} 
    \startdata
    \hline
    \hline
    Baseline scenario & $2\times10^{5}$\\
    \hline
    \citet{Peacock2019a} Model 1A SED (Fig.~\ref{fig:SED}) & $3\times10^{4}$\\
    $T_{surf}=250$ K & $2\times10^{5}$ \\
    \texttt{Photochem} RCE T-P profile; $P_{H_{2}O, surf}=1.4\times10^{-6}$ bar  (Fig.~\ref{fig:TP}) & $2\times10^{5}$ \\
    \tablenotemark{a}Max. oxidant deposition:
 $v_{dep, CO}=1\times10^{-4}$ cm$^{-2}$ s$^{-1}$, $v_{dep, O_{2}}=1\times10^{-4}$ cm$^{-2}$ s$^{-1}$, $v_{dep, CO_{2}}=4\times10^{-3}$ cm$^{-2}$ s$^{-1}$ & $2\times10^{5}$ \\    
    \tablenotemark{b}Max. haze formation: $r_d=0.1~\mu$m, $v_{dep, C_{2}H_{6}}=v_{dep, C_{2}H_{4}}=v_{dep, C_{2}H_{2}}=v_{dep, CH_{3}CN}=v_{dep, HCCCN}=v_{dep, HCN}=0$ & $2\times10^{5}$ \\    
    $K_{ZZ}=10^8$ cm$^{2}$ s$^{-1}$ & $6\times10^{4}$\\    
    $K_{ZZ}=10^4$ cm$^{2}$ s$^{-1}$ & $2\times10^{5}$\\ 
    $P_{surf}=0.15$ bar & $5\times10^{4}$\\    
    $P_{surf}=15$ bar & $2\times10^{5}$\\    
    \enddata
    \tablenotetext{a}{Conditions which minimize buildup of oxidized species from H$_2$O photolysis. The adopted deposition velocities correspond to piston-limited CO$_2$/CO/O$_2$ deposition velocities \citep{Harman2015} for an ocean-covered TRAPPIST-1e.}
    \tablenotetext{b}{Conditions which maximize haze formation. We have decreased $r_d$ from 1 $\mu$m to 0.1 $\mu$m, which decreases the settling velocity and therefore increases atmospheric haze lifetime, and we have eliminated the dry deposition of haze species.}
\end{deluxetable}


We also find that $\tau_{\text{CH}_{4}}$ increases as p$_{\text{CH}_{4}}$ increases\footnote{We recovered qualitatively similar behavior in the {\tt MEAC} photochemical model \citep{Hu2012, Ranjan2022b}. This does not constitute strong independent confirmation since {\tt MEAC} does not include organic haze formation which occurs in this regime, but still modestly increases our confidence in this finding since photochemical runaway is independent of hazes \citep{Kasting1990, Ranjan2022b}}. We explain this phenomenon as another example of photochemical runaway \citep{Kasting1990, Segura2005, Ranjan2022b}: CH$_4$ loss is ultimately driven by UV photolysis \citep{Wilson2009}, either directly: 
\begin{align}
    CH_4 + h\nu \rightarrow \text{products}
\end{align}
or indirectly, e.g.:
\begin{align}
\begin{split}
        C_2H_2 + h\nu &\rightarrow C_2H + H\\
        C_2H+CH_4 &\rightarrow C_2H_2 + CH_3\\
        \hline
        \text{Net: } CH_4 &\rightarrow CH_3 + H
\end{split}
\end{align}
The supply of UV photons capable of dissociating CH$_4$ is finite. \citet{Lora2018} find direct photolysis to dominate CH$_4$ loss around M-dwarf stars due to their emission of more of their UV output in Lyman-$\alpha$. CH$_4$ Lyman-$\alpha$ absorption saturates for p$_{\text{CH}_{4}}>2\times10^{-9}$ bar, meaning that above this threshold, the CH$_4$ photolysis rate generally increases sublinearly with p$_{\text{CH}_{4}}$. Consequently, to first order, larger p$_{\text{CH}_{4}}$ leads to a larger $\tau_{\text{CH}_{4}}$. Conversely, smaller p$_{\text{CH}_{4}}$ decreases $\tau_{\text{CH}_{4}}$; for example, $\tau_{\text{CH}_{4}}(\text{pCH}_{4}=3\times10^{-6}~\text{bar})=2\times10^{3}$ years, a geological eye-blink. This means that smaller inferred values for p$_{\text{CH}_{4}}$ are even less plausible than our baseline p$_{\text{CH}_{4}}=0.03$ bar. 


At the very highest p$_{\text{CH}_{4}}$ we consider here (p$_{\text{CH}_{4}}\geq0.03$ bar), the CH$_4$ lifetime exceeds the upper limit set by assuming the diffusion-limited escape of total hydrogen controls the CH$_4$ lifetime (\citealt{Thompson2022}; Figure~\ref{fig:LCH4_PCH4}). This upper limit is based on the fact that escape of hydrogen makes the reaction $CH_4 + h\nu  \rightarrow CH_3+ H$ irreversible, and therefore that the total hydrogen escape rate is a lower limit on the CH$_4$ loss rate \footnote{\citet{Thompson2022} use the equivalent H$_2$ escape rate to set CH$_4$ loss rate, but since the loss of a single H is enough to irreversibly destroy a molecule of CH$_4$, we instead use the equivalent H escape rate, i.e. twice the H$_2$ rate.}:
\begin{align}
    L_{\text{CH}_{4}} &\geq \Phi_{H, tot}^{\uparrow}\\
               &\geq \bigl( (r_{H}+2r_{H_{2}} + 2r_{H_{2}O} + 4r_{\text{CH}_{4}})|_{\text{lower stratosphere}}\bigl) \bigl( \frac{b_{H-H_{2},N_{2}}}{H_{atm}}|_{\text{homopause}}\bigl) \\
               &> \bigl(4r_{\text{CH}_{4}})|_{\text{lower stratosphere}}\bigl) \bigl(\frac{b_{H-H_{2},N_{2}}}{H_{atm}}|_{\text{homopause}} \bigl)
\end{align}
\noindent where $r_X$ is the molar concentration of species $X$, $H_{atm}$ is the atmospheric scale height, and $b_{H-H_{2},N_{2}}$ is a composite binary diffusion parameter for both H and H$_2$, estimated following \citet{CatlingKasting2017}, using underlying binary diffusion parameters for H and H$_2$ through N$_2$ from \citet{Banks1973}. However, this approach assumes that there are sufficient UV photons available to convert CH$_4$ to H and H$_2$ in the upper atmosphere. For p$_{\text{CH}_{4}}\geq0.03$ bar in this planetary scenario, this assumption begins to fail: CH$_4$ cannot convert to H fast enough to keep up with escape. We attribute this to the overall lower total UV output ($100-300$ nm) of M-dwarf stars compared to Sun-like stars \citep{Segura2005}. As a consequence, $\tau_{\text{CH}_{4}}$ can be underestimated by up to an order of magnitude using the diffusion-limited escape of total hydrogen approximation, compared to the estimate using a full photochemical model (Figure~\ref{fig:LCH4_PCH4}). Despite this underestimation, $\tau_{\text{CH}_{4}}$ remains $\gtrsim2$ orders of magnitude shorter for a warm M-dwarf exo-Titan compared to true Titan. 

We also estimated $\tau_{\text{CH}_{4}}$ assuming $\tau_{\text{CH}_{4}}\propto\frac{1}{F_{EUV}}$, where ${F_{EUV}}$ is the extreme-ultraviolet (EUV) flux, as proposed by \citet{Turbet2018}. We assume the EUV to correspond to 12.4-91.2 nm following \cite{Wheatley2017}, cited by \citet{Turbet2018}, and estimate $F_{EUV}$ at TRAPPIST-1e from the \citet{Wilson2021} semi-empirical SED (Figure~\ref{fig:SED}). We find this approach to underestimate $\tau_{\text{CH}_{4}}$ by about two orders of magnitude relative to the full photochemical calculation (Figure~\ref{fig:LCH4_PCH4}), indicating the utility of considering detailed photochemistry when attempting to quantify the likelihood of exo-Titan scenarios. 

\subsection{(Im)Plausibility of Warm Exo-Titans}
Our results suggest that \textit{a priori}, warm exo-Titans orbiting M-dwarfs are unlikely. Across the range of p$_{\text{CH}_{4}}$ we consider here, $\tau_{\text{CH}_{4}}$ on a warm M-dwarf exo-Titan is $\leq2\times10^{-2}\times\tau_{\text{CH}_{4}, \text{true Titan}}$, with lower p$_{\text{CH}_{4}}$ corresponding to lower $\tau_{\text{CH}_{4}}$. This means that all else being equal, detecting a CH$_4$-rich warm M-dwarf exo-Titan is $\leq 0.02\times$ as likely as detecting CH$_4$-rich true Titan. The CH$_4$-rich modern Titan may itself be transient \citep{Lorenz1997, Tobie2006, Wong2015}. For example, in the model of \citet{Tobie2006}, about 3.2 Gyr of the 4.55 Gyr history of Titan corresponds to a high-outgassing, CH$_4$ rich phase, and imaging Titan outside of these epochs would yield a thin, cold, N$_2$-dominated atmosphere similar to Pluto's \citep{Wong2015}. This corresponds to a probability of $0.7$ of imaging a CH$_4$-rich Titan, which would imply the probability of imaging a warm exo-Titan is $\leq0.01$, all else being equal. 

All else may not be equal. The methane inventories on warm exo-Titans may be larger than on true Titan due to larger size, increasing the cumulative CH$_4$ lifetime (and therefore detection probability) by up to an order of magnitude. The planet-integrated CH$_4$ loss rate scales with the surface area ($\propto R_P^2$), while the planetary CH$_4$ inventory scales with mass ($\propto w_{\text{CH}_{4}}\rho_{P} R_P^3$), where $R_P$ is the planetary radius, $\rho_P$ is the planetary mass density, and $w_{\text{CH}_{4}}$ is the CH$_4$ mass fraction. Therefore, the CH$_4$ lifetime should scale as $w_{\text{CH}_{4}}\rho_P R_P$. The radius of Titan $R_{Titan}=0.4R_\oplus$ \citep{CatlingKasting2017}, suggesting that Earth-sized exo-Titans can increase $\tau_{\text{CH}_{4}}$ by a factor of 2.5 due to increased size. Additionally, thermodynamic modeling suggests that under the conditions found in the interior of volatile-rich super-Earths, high-pressure methane-filled ices (Ice Ih) should be stable with $\approx2\times$ higher CH$_4$:H$_2$O ratios compared to methane clathrates \citep{Levi2013, Levi2014, Levi2019, Levi2023}, suggesting  that $w_{\text{CH}_{4}}$ may be up to $2\times$ on a super-Earth exo-Titan compared to true Titan. Lastly, $\rho_P$ increases with $R_P$ due to compression and is $1.3\times$ larger for a 1-$R_\oplus$ Earth-composition body compared to a 0.4-$R_\oplus$ Earth-composition\footnote{\url{https://lweb.cfa.harvard.edu/~lzeng/tables/massradiusEarthlikeRocky.txt}, accessed 08/04/2025} \citep{Zeng2019}. Taken together, a 1 $R_\oplus$ exo-Titan may have a CH$_4$ lifetime enhanced by up to $7\times$ compared to a Titan-sized exoplanet. However, even these generous assumptions still leave warm exo-Titans with detection probabilities of $<0.1\times$ true Titan. 

If methane is released to warm exo-Titan atmospheres in infrequent, large-volume events as proposed for Titan \citep{Tobie2006}, exo-Titans with small inferred p$_{\text{CH}_{4}}$ are less plausible than exo-Titans with high inferred pCH${_{4}}$. This is because $\tau_{\text{CH}_{4}}$ decreases as p$_{\text{CH}_{4}}$ decreases, meaning it is less likely to image an exo-Titan as it is passing through the short-lived low p$_{\text{CH}_{4}}$ phases compared to the longer-lived high p$_{\text{CH}_{4}}$ phases. However, it is important to note that this finding is restricted to p$_{\text{CH}_{4}} \leq 0.1$ bar. For p$_{\text{CH}_{4}}\gtrsim0.1-1$ bar, the temperature increases with p$_{\text{CH}_{4}}$ \citep{Turbet2018}, triggering a positive feedback loop that would result in a hot, thick CH$_4$ atmosphere \citep{Levi2014}. The thermal stability of CH$_4$ decreases as temperature increases (Appendix~\ref{sec:ch4thermo}), meaning there is a possibility that for a massive enough CH$_4$ atmosphere, thermal loss would emerge as major CH$_4$ loss mechanism, increasing $L_{\text{CH}_{4}}$ and decreasing $\tau_{\text{CH}_{4}}$. Coupled climate-photochemistry modeling with a climate model validated for CH$_4$-dominated atmospheres is required to resolve this scenario. 

\citet{Levi2014} propose that instead of catastrophic release, CH$_4$ on exo-Titan super-Earths may instead be released slowly but continuously ($\leq10^{10}$ cm$^{-2}$ s$^{-1}$) to the atmosphere via convective overturn in the mantle due to the different dynamics of methane-filled Ice Ih, which is expected to be the dominant form of water at the high pressures in super-Earth exo-Titan mantles \citep{Levi2019, Levi2023, Journaux2020}. \citet{Thompson2022} argue that such atmospheres are unstable, because melting liberates CH$_4$ which heats the planet, triggering more melting in a positive feedback loop. However, \citet{Thompson2022} neglected the previously noted strong antigreenhouse effect of CH$_4$ at low p$_{\text{CH}_{4}}$ \citep{RamirezKaltenegger2018, Turbet2018}. This antigreenhouse effect means that if starting from p$_{\text{CH}_{4}}<0.1$ bar, increasing p$_{\text{CH}_{4}}$ cools the surface \citep{Turbet2018}, constituting a negative feedback regulating CH$_4$ outgassing \citep{Levi2014} to the rate required to compensate for photochemical loss. At the low rates ($\leq10^{10}$ cm$^{-2}$ s$^{-1}$) proposed for this mechanism, p$_{\text{CH}_{4}}$ is also low (p$_{\text{CH}_{4}}<3\times10^{-5}$ bar; Figure~\ref{fig:LCH4_PCH4}). Detection of a warm exo-Titan with low p$_{\text{CH}_{4}}$ would constitute evidence in favor of the \citet{Levi2014} continuous outgassing scenario, as opposed to the catastrophic release scenario. 

We have focused our modeling on TRAPPIST-1e in this paper, but we expect our findings generalize to warm exo-Titans in general. Our basic finding that warm exo-Titan atmospheres have orders of magnitude lower $\tau_{\text{CH}_{4}}$ compared to true Titan is driven by the higher CH$_4$-dissociating FUV flux ($\leq152$ nm), not by planetary mass/radius, thermal structure, or vertical convection rate. Two orders of magnitude of FUV flux increase compared to Titan come from the specification that the planet be warm ($T_{surf}\geq200$ K), and the higher activity of M-dwarf stars can further enhance the FUV irradiation incident on warm exo-Titans. To test this expectation, we conducted a sensitivity test wherein we applied SEDs corresponding to a representative sample of M-dwarfs\footnote{All SEDs except LHS-2686 for which our model did not converge.} drawn from \cite{Wilson2025} to our p$_{\text{CH}_{4}}=3\times10^{-2}$ bar scenario, and found comparable $\tau_{\text{CH}_{4}}$ to the TRAPPIST-1e irradiation case. We conclude that the implausibility of warm exo-Titans is not specific to TRAPPIST-1. 

Finally, we note that photochemical destruction is not the only barrier to the existence of CH$_4$ on warm exo-Titans. CH$_4$ is a light molecule, and may escape hydrodynamically on true Titan \citep{Strobel2008, Strobel2009}, though this is contested (e.g., \citealt{Bell2014}). While escape of CH$_4$ is mild and subdominant to photolysis as a loss channel on Titan \citep{Yelle2008}, the situation may be different on M-dwarf planets like TRAPPIST-1e, which are subject to much higher XUV irradiation levels (e.g., \citealt{Nakayama2022, vanLooveren2024, Chatterjee2024}), and which therefore may conceivably experience vigorous hydrodynamic escape of CH$_4$ at rates comparable to or in excess of photolytic loss. Hydrodynamic escape modeling of M-dwarf super-Earths with high-$\mu$ atmospheres including relevant atomic and molecular line cooling is required to address this possibility \citep{Chatterjee2024}. In the interim, the warm exo-Titan detection probabilities we provide here must be considered upper limits.

\subsection{Possible CO$_\text{X}$ in Warm Exo-Titan Atmospheres \label{sec:cox}}
\begin{figure}[h]
\centering

\begin{subfigure}{0.6\linewidth}
    \includegraphics[width=1\linewidth]{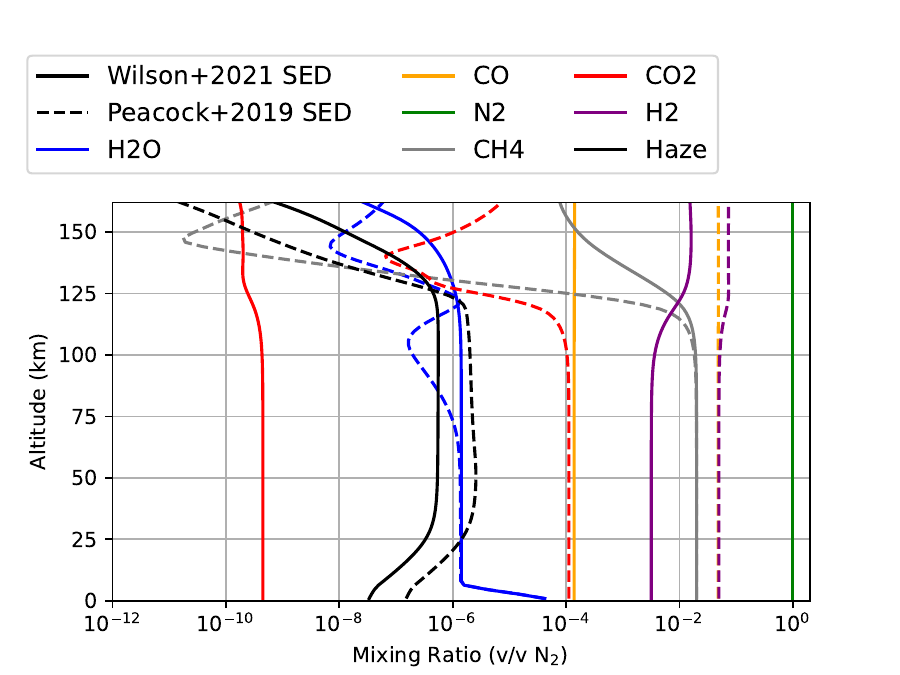}
    \caption{Volume mixing ratios with respect to N$_2$ for CHO and haze species for p$_{\text{CH}_{4}}=3\times10^{-2}$ bar, forced by the \citet{Wilson2021} (solid line) and \citet{Peacock2019a} (dashed line) SEDs. CO$_\text{X}$ is formed in both cases, but its abundance is very sensitive to the choice of SED. Specifically, the \citet{Peacock2019a} SED leads to much higher predicted CO$_\text{X}$, which we attribute to greater production of the oxidant OH due to increased H$_2$O photolysis driven by the higher NUV irradiation in this SED.} 
\end{subfigure}

\begin{subfigure}{\linewidth}
    \centering
    \begin{subfigure}{0.49\linewidth}
        \includegraphics[width=1\linewidth]{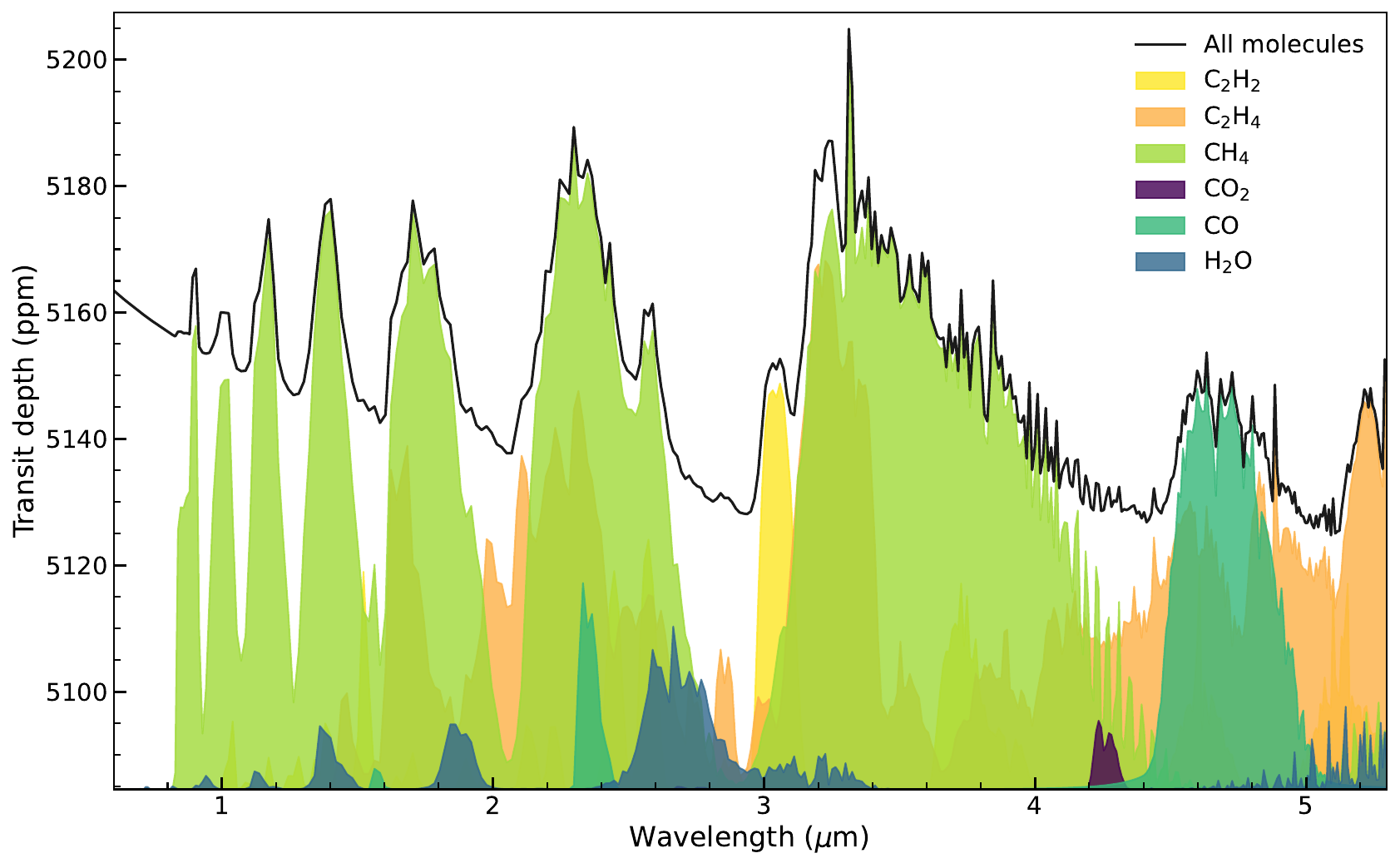}
    \end{subfigure}
    \hfill
    \begin{subfigure}{0.49\linewidth}
        \includegraphics[width=1\linewidth]{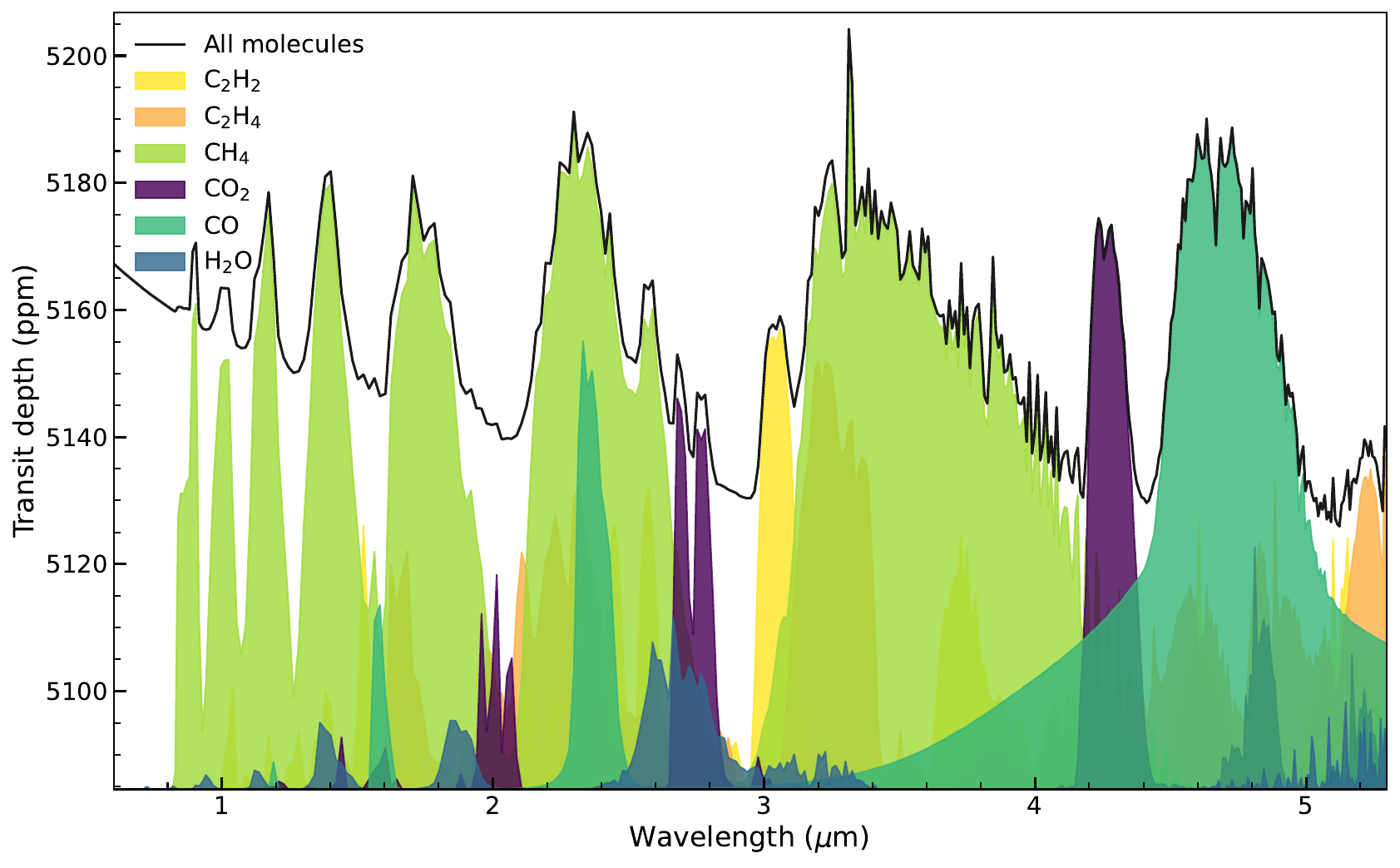}
    \end{subfigure}
    \caption{Simulated transmission spectrum for warm exo-Titan TRAPPIST-1e (p$_{\text{CH}_{4}}=3\times10^{-2}$ bar) forced by \citet{Wilson2021} SED on the left and forced by \citet{Peacock2019a} SED on the right. Photochemical CO but not CO$_2$ generated by CH$_4$ photoxidation may be detectable for the \citet{Wilson2021} SED, whereas both CO and CO$_2$ may be detectable for the \citet{Peacock2019a} SED.}
\end{subfigure}

\caption{Atmospheric CO$_\text{X}$ on a warm exo-Titan TRAPPIST-1e (p$_{\text{CH}_{4}}=3\times10^{-2}$ bar). The simultaneous presence of abundant CH$_4$ and H$_2$O in the warm exo-Titan scenario results in abundant CO and maybe CO$_2$, which are potentially detectable.}
\label{fig:COx}
\end{figure}

Warm exo-Titans may have abundant oxidized carbon (CO, CO$_2$; CO$_\text{X}$) in their atmospheres due to their higher temperatures (Figure~\ref{fig:COx}). Titan has very little CO$_\text{X}$ because it is so exceptionally cold ($T_{surf}=94$ K) that negligible gaseous H$_2$O is available to oxidize CH$_4$ to CO$_\text{X}$ (additionally, CO$_2$ condenses). By contrast, on warm exo-Titans with T$_{surf}\geq200$ K, H$_2$O is available as an oxidant, meaning that photochemical oxidation of CH$_4$ to CO$_\text{X}$ is a possibility. This possibility has precedent in the literature: \citet{Zahnle2020} model the photochemistry of a warm Titan-like CH$_4$-N$_2$ atmosphere on early Earth. They find that for stratospheric [H$_2$O]$\leq0.1$ ppm, CH$_4$ mostly converts to hydrocarbon haze as on Titan, whereas for stratospheric [H$_2$O]$\geq 1$ ppm, CH$_4$ instead mostly oxidizes to CO$_\text{X}$. The stratospheric H$_2$O concentration on temperate worlds with water is set by the stratospheric temperature through its control on the saturation vapor pressure of water \citep{Wordsworth2013CO2}. For N$_2$-CH$_4$ atmospheres orbiting M-dwarf stars, the stratospheres are likely to be warm and therefore moist due to the strong anti-greenhouse effect (e.g., Figure~\ref{fig:TP}; \citealt{RamirezKaltenegger2018}), implying efficient CO$_\text{X}$ production. 

We tested this possibility in our model. We found that appreciable CO$_\text{X}$ was indeed produced, but its abundance was sensitive to choice of SED (Figure~\ref{fig:COx}). For p$_{\text{CH}_{4}}=3\times10^{-2}$ bar and the \citet{Wilson2021} semi-empirical SED, CO volume mixing ratios with respect to N$_2$ are at the $10^{-4}$ level, at which they generate potentially-detectable spectral features (Figure~\ref{fig:COx}). For context, Titan has CO volume mixing ratios of 10-100 ppm \citep{Horst2008}, and it is sensible for warm exo-Titan CO mixing ratios to be comparable or greater because of the greater availability of H$_2$O as oxidant. For the \citet{Peacock2019a} model 1A SED, NUV emission is higher, driving more intense H$_2$O photolysis and increasing CO and CO$_\text{X}$ concentrations by 2 and 5 orders of magnitude, respectively compared to the \citet{Wilson2021} SED. At such concentrations, CO$_2$ may also be detectable (Figure~\ref{fig:COx}). Detection of CO$_\text{X}$ would therefore corroborate an exo-Titan scenario (Section~\ref{sec:obs}).

In addition to photochemical production, CO$_\text{X}$ on warm exo-Titans should be directly outgassed along with CH$_4$. The CO$_2$:CH$_4$ mass ratio on Titan is 11-16:1 \citep{Thompson2022, Tobie2012}, implying that CO$_2$ should be released to the atmosphere in comparable quantities to CH$_4$. True Titan is cold enough that released CO$_2$ freezes out, but this is not applicable to warm exo-Titans. On Earth outgassed CO$_2$ is removed from the atmosphere by weathering, but this mechanism would not be applicable to a warm exo-Titan with a thick submarine ice layer sealing off access to the rock which is the source of cations for carbonate precipitation. Instead, pCO$_2$ should be regulated by formation of CO$_2$ clathrate, which is much less efficient than chemical weathering and allows for much larger pCO$_2$ \citep{Levi2017, Ramirez2018}. In summary, it is likely for a warm exo-Titan to feature CO$_\text{X}$ in its atmosphere, and lack of such corroborating CO$_\text{X}$ reduces the plausibility of an exo-Titan interpretation to a given spectrum.


\subsection{Implications for JWST Observations \label{sec:obs}}
Because warm exo-Titans are unlikely, a high standard of proof is required to claim one's detection. Our work suggests that given our current understanding how a Titan-like atmosphere forms and photochemically evolves, the probability of detecting a warm exo-Titan TRAPPIST-1e is $<0.1\times$ the probability of detecting Titan, and possibly much lower, depending on the putative p$_{\text{CH}_{4}}$. It is worth bearing in mind that Titan's current CH$_4$-rich state is likely itself transient \citep{Wong2015}, reducing further the absolute probability of detecting a warm exo-Titan. This is relevant because in practice, remote detection of molecular features in planetary atmospheres follows a Bayesian approach: a higher standard of proof is demanded for features which are judged theoretically likely compared to those that are judged theoretically unlikely \citep{Zahnle2011, Lincowski2018, Wogan2024a, Welbanks2025}.  Our constraints on absolute probability of detecting a warm exo-Titan provide a quantitative prior in Bayesian inference of putative exo-Titans \citep{Catling2018}. A corollary of our study is the prediction that methane planets \citep{Helled2015} are unlikely on photochemical grounds. This prediction has begun to receive observational confirmation, with CH$_4$ atmospheres ruled out on multiple warm super-Earths by JWST observations \citep{Lustig-Yaeger2023, Scarsdale2024}.

Our findings suggest that a high standard of proof should apply to claiming a CH$_4$ detection on temperate planets, because warm exo-Titans are, within the limits of current knowledge, \textit{a priori} unlikely and therefore demand strong evidence. In particular, to promote a candidate CH$_4$ signal to a detection claim, we advocate requiring: 
\begin{enumerate}
    \item \textit{Signal authenticity.} We recommend verifying that a putative detection is robust to data detrending technique \citep[e.g., ][]{Gibson2011a}.
    \item \textit{Signal attribution.} We recommend concurrently constraining stellar contamination alongside the putative planetary transmission spectrum, to determine whether the putative spectral features can be attributed to the planet versus stellar activity \citep[e.g.,][]{Espinoza2025}.
    \item \textit{Signal uniqueness.} We recommend detection of multiple CH$_4$ spectral features \citep[e.g., ][]{Madhusudhan2023}, to increase the confidence that other molecules cannot plausibly explain the signal \citep{Niraula2025}. 
    \item \textit{Signal consensus.} We recommend the detection be recovered using multiple independent data reduction and retrieval tools \citep[e.g., ][]{Schmidt2025}, to mitigate the risk that the detection is an artifact of assumptions or errors in a specific code. 
\end{enumerate}
\noindent A similarly high standard of proof should apply to other candidate warm exo-Titans (e.g., GJ1132b; \citealt{Swain2021, Adams2022, Mugnai2021, Libby-Roberts2022, MayMacDonald2023}). 

Our findings have implications for JWST observations. Specifically, retrievals applied to the recently-reported JWST GTO TRAPPIST-1e transmission spectrum \citep{Espinoza2025} weakly favor a CH$_4$-rich, CO$_2$-poor, high-$\mu$ atmosphere \citep{Glidden2025}. \citet{Glidden2025} used analogies to Solar System bodies to contextualize the atmospheric scenarios allowed by the data. While they did not find a direct Solar System analog, they noted that the favored atmospheric scenario was most comparable to a warm Titan, with other possible explanations for the signal being stellar activity or random noise. Here, we have performed a deeper investigation to determine the plausibility of the warm exo-Titan scenario through detailed photochemical modeling, which reveals that a warm exo-Titan TRAPPIST-1e is \textit{a priori} unlikely due to the short CH$_4$ lifetime. Consequently, our Bayesian expectation is that follow-up observations of TRAPPIST-1~e will reveal the tentative inference of abundant CH$_4$ to be explained by stellar contamination or noise. Such observations are now underway (JWST-GO-06456; PIs N. Espinoza, N. Allen).




If future JWST observations (e.g., of TRAPPIST-1~e; JWST-GO-06456) reveal firm detections of warm exo-Titans, falsifying our theoretical expectations, then our theoretical understanding of exo-Titans is likely incomplete or incorrect and needs to be revisited. The largest keyhole for CH$_4$-rich atmospheres comes from our limited understanding of the interiors of exo-Titans \citep{Levi2013, Levi2014,Levi2019}. Estimates of exo-Titan CH$_4$ inventory here and in other works (e.g., \citealt{Thompson2022}) is based on simple scaling of true Titan, and does not consider detailed modeling of the interior structure and CH$_4$ storage and transport on a more massive waterworld. Consideration of these detailed effects may reveal that exo-Titans can store and slowly outgas larger volumes of CH$_4$ than considered by our analysis. Aside from the interior, it is also possible that our understanding of CH$_4$ photochemistry is incorrect. In particular, we have not treated in detail the regime where p$_{\text{CH}_{4}}>0.15$ bar, because we expect dramatic changes in climate and interior-atmosphere interactions in this regime which our modeling framework has not been validated for. It is possible that in this regime $\tau_{\text{CH}_{4}}$ may be high, though our expectation is the opposite on thermochemical grounds (Appendix~\ref{sec:ch4thermo}). If both interior and photochemical explanations fail and the observational inference of abundant CH$_4$ remains secure, it may be necessary to consider alternative explanations (e.g., \citealt{Thompson2022}). 

\section{Conclusions}
We have used an atmospheric photochemistry model to evaluate CH$_4$ lifetimes $\tau_{\text{CH}_{4}}$ on warm exo-Titans (exoplanets with N$_2$-CH$_4$ atmospheres overlying rock-ice planets with volatile mass fractions $>10\%$). We find that warm Exo-Titans have cumulative CH$_4$ lifetimes $\leq 0.02\times$ true Titan (i.e., $\leq 400,000$ years), rising $\leq 0.1\times$ true Titan under very generous assumptions as to CH$_4$ content. This implies the absolute probability of detection a warm exo-Titan is $<0.1$, and likely $<0.01$, providing a quantitative prior that can be used in Bayesian analyses of putative exo-Titans. This finding is driven by UV instellation, and is not sensitive to temperature structure, eddy diffusion, or details of the haze formation scheme. These detection probabilities must be considered upper limits because atmospheric escape may further shorten $\tau_{\text{CH}_{4}}$. 

Since exo-Titans are \emph{a priori} unlikely, a high standard of proof is required to claim one's detection. Specifically, we argue that to claim an exo-Titan, (1) the detection should be insensitive to data detrending technique, (2) stellar contamination must be constrained alongside the planetary spectrum, (3) multiple bands of CH$_4$ must be detected, and (4) molecular features must be robust to choice of data reduction pipeline and retrieval/forward model tool. Detection of oxidized carbon species (CO, CO$_\text{X}$) expected from photochemistry or outgassing would corroborate an exo-Titan interpretation to a putative JWST transmission spectrum. As a corollary, our work explains the nondetection of methane planets \citep{Helled2015} to date \citep{Lustig-Yaeger2023, Scarsdale2024}.

Our work shows the utility of using a full photochemical model to estimate CH$_4$ lifetime in the exo-Titan regime. Simpler analytic approaches to estimating $\tau_{\text{CH}_{4}}$ (e.g., scalings based on $F_{EUV}$ or diffusion-limited escape of equivalent H$_2$) can underestimate $\tau_{\text{CH}_{4}}$ by 1-2 orders of magnitude. A numerical 1D photochemical model is required to capture the nonlinearities inherent to atmospheric photochemistry.

\begin{acknowledgments}
We thank N. Espinoza for many insightful discussions and comments. This research has made use of NASA’s Astrophysics Data System. This paper reports work carried out in the context of the JWST Telescope Scientist Team (https://www.stsci.edu/$\sim$marel/jwsttelsciteam.html, PI: M. Mountain). Funding is provided to the team by NASA through grant 80NSSC20K0586. S.R. thanks the Space Telescope Science Institute for support via grant JWST-GO-06456.018-A. N.F.W was supported by the NASA Postdoctoral Program.

Software required to reproduce calculations shown in this paper are available on Github (https://github.com/sukritranjan/ExoTitan-ForGit) and at \citet{ranjan_2025_exotitan_zenodo}.

\end{acknowledgments}

\begin{contribution}

SR conceived and executed the research and drafted the manuscript. NFW designed the Titan and exo-Titan templates, consulted on deployment of Photochem, and checked and verified convergence of SR's model calculations. AG executed spectral forward models and retrievals. JW conducted MEAC modeling to confirm increase in $\tau_{\text{CH}_{4}}$ with p$_{\text{CH}_{4}}$ and wrote thermodynamics script. KBS and NK advised on observational standards of proof. TK advised on Titan. All authors reviewed and commented on the manuscript. 

\end{contribution}

%
\facilities{JWST}

\software{MEAC \citep{Hu2012, Ranjan2022b},
          Photochem \citep{Wogan2023, Wogan2024_photochem}, 
          petitRADTRANS \citep{Molliere2019prt}}


\appendix
\section{Thermochemical Stability of CH$_4$ \label{sec:ch4thermo}}
\begin{figure}[h]
\centering
\includegraphics[width=0.7\linewidth]{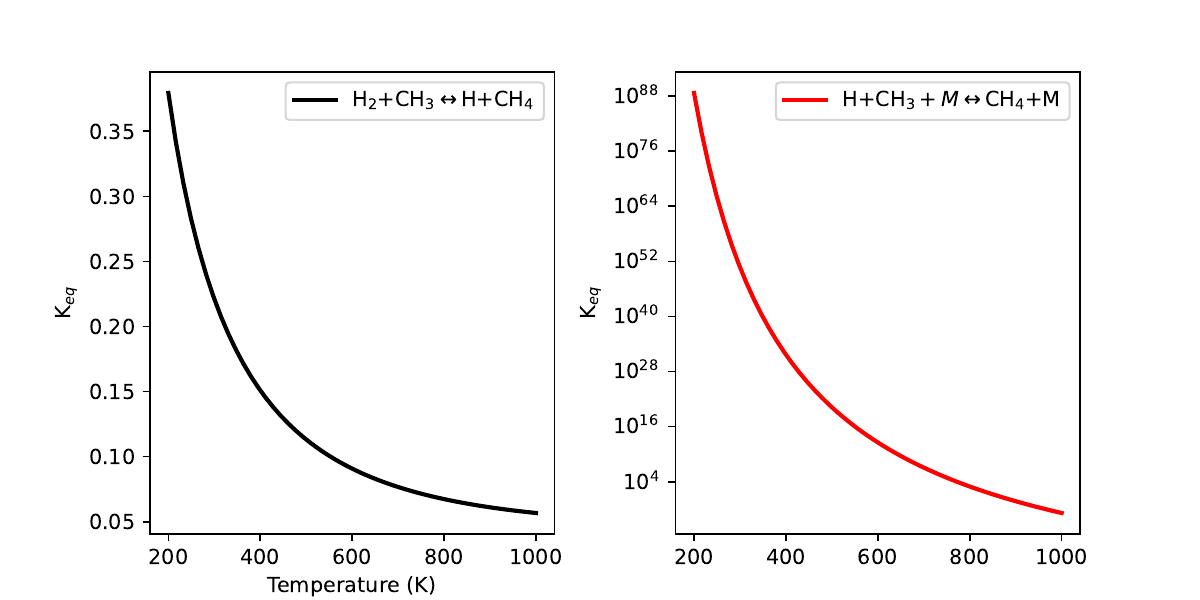}
\caption{Equilibrium constant for the key reactions $H_2+CH_3\leftrightarrow H + CH_4$, $H+CH_3+M\leftrightarrow CH_4+M$. We find $K_{eq}$ decreases with temperature for both reactions, implying reduced CH$_4$ stability at higher temperatures.} 
\label{fig:Keq}
\end{figure}

In this Appendix, we consider whether CH$_4$ is likely to be thermochemically stable at high temperatures, as might prevail on an exo-Titan with p$_{\text{CH}_{4}}>1$ bar, for which CH$_4$ liberated by melting surface ice heats the surface melting more ice in a positive feedback loop \citep{Turbet2018, Thompson2022}. To explore this question, we calculate the equilibrium constant 
$K_{eq}$ for the key reactions
\begin{align}
    H_2+CH_3 &\leftrightarrow H + CH_4\\
    H+CH_3+M &\leftrightarrow CH_4+M
\end{align}
\noindent identified as stabilizing CH$_4$ in hot exo-Titans ($T_{surf}=950$ K, $T_{strat}=480$ K) by \citet{Adams2022}. We calculate $K_{eq}$ following the prescription in \citet{Rimmer2016}, using thermodynamic parameters from \citet{Burcat2005}, accessed via \url{https://respecth.elte.hu/}. We find (Figure~\ref{fig:Keq}) that $K_{eq}$ for both reactions drastically decreases as $T$ increases, meaning that it is possible that in addition to being photochemically unstable, at high $T$ CH$_4$ may also degrade thermally. Therefore, the photochemical estimate for $\tau_{\text{CH}_{4}}$ may overestimate the true $\tau_{\text{CH}_{4}}$ for hot worlds. 

\bibliography{main.bib}{}
\bibliographystyle{aasjournalv7}



\end{document}